\documentclass[fleqn,twoside]{article}
\usepackage{espcrc2,epsfig}

\title{Status of the QCDOC project\thanks{Presented
    by NHC and TW at Lattice 2001, Berlin.}}

\author{
P.A.~Boyle\address[Ed]{Department of Physics and Astronomy, University 
  of Edinburgh, Edinburgh EH9 3JZ, Scotland}\address[Columbia]{Department 
  of Physics, Columbia University, New York, NY, 10027},
D.~Chen\address[IBM]{IBM T.J.~Watson Research Center, Yorktown
  Heights, NY, 10598}, 
N.H.~Christ\addressmark[Columbia], 
C.~Cristian\addressmark[Columbia], 
Z.~Dong\addressmark[Columbia],
A.~Gara\addressmark[IBM],
B.~Jo\'o\addressmark[Ed]\addressmark[Columbia],
C.~Kim\addressmark[Columbia], 
L.~Levkova\addressmark[Columbia], 
X.~Liao\addressmark[Columbia],
G.~Liu\addressmark[Columbia],
R.D.~Mawhinney\addressmark[Columbia],
S.~Ohta\address{Institute for Particle and Nuclear Studies,
  KEK, Tsukuba, Ibaraki, 305-0801, Japan}\address[RBRC]{RIKEN-BNL 
  Research Center, Brookhaven National Laboratory, Upton, NY, 11973},
T.~Wettig\addressmark[RBRC]\address{Department of Physics, Yale
  University, New Haven, CT, 06520-8120},
A.~Yamaguchi\addressmark[Columbia]}
     
\begin{document}

\begin{abstract}
  A status report is given of the QCDOC project, a massively parallel
  computer optimized for lattice QCD using system-on-a-chip
  technology.  We describe several of the hardware and software features
  unique to the QCDOC architecture and present performance figures
  obtained from simulating the current VHDL design of the QCDOC chip 
  with single-cycle accuracy.
  \vspace{1pc}
\end{abstract}

\maketitle

\section{INTRODUCTION}

The debate over which computing platforms are suitable for the
simulation of lattice gauge theories has continued at this year's
lattice conference.  It has been shown that recent off-the-shelf PC
processors by Intel or AMD are capable of delivering impressive
floating-point performance on the order of 1 GFlops if the vector
processing units are employed and cache management is optimized
\cite{Luescher01}.  Unfortunately, this single-node performance cannot
be fully utilized in a PC cluster if the local lattice volume becomes
small.  In this case, the surface-to-volume ratio is large, and the
communications latency inherent in standard network solutions such as
Ethernet or Myrinet slows down the individual processors.  Therefore,
to achieve reasonable efficiencies with a PC cluster, the local
lattice volume must not be too small.  This implies that for a given
total lattice volume, the number of nodes working on a single problem
is limited.

However, to finish a dynamical fermion calculation on a moderately
large lattice in a reasonable amount of time, it is essential to
distribute the total volume onto as many nodes as possible, which
implies very small local lattice volumes, perhaps as small as $2^4$.
This requires communications between neighboring nodes with extremely
low latencies that cannot be achieved using off-the-shelf networking
components.  Massively parallel machines with custom-designed
communications hardware (with latencies 10--100 times smaller than
in the case of Myrinet) appear to be the only viable alternative.
This is especially true in the field of lattice gauge theory with
its very regular communications requirements.  In this contribution,
we describe the design of the QCDOC architecture which is capable of
delivering computing power in the tens of TFlops range at a
price/performance ratio of 1~US-\$ per sustained MFlops.

\section{OVERVIEW OF QCDOC}

A first description of the QCDOC project has been given at last year's
lattice conference \cite{Chen01}.  In the nomenclature of parallel
computers, this is a multiple-instruction, multiple-data (MIMD) 
machine with distributed memory.  The name
QCDOC stands for ``QCD On a Chip''.  An essential element of this
project is the collaboration with IBM Research and the resulting
ability to utilize IBM's system-on-a-chip technology.  Nowadays it is
possible to integrate most of the components that make up a single
processing node on a single chip, creating an application-specific
integrated circuit, or ASIC.  The QCDOC chip is such an ASIC, consisting of
\begin{itemize}\itemsep -1mm
\item 500 MHz, 32-bit PowerPC 440 processor core
\item 64-bit, 1 GFlops floating-point unit
\item 4 MBytes on-chip memory (embedded DRAM)
\item controllers for embedded and external memory
\item nearest-neighbor serial communications unit with estimated latencies of 
  120ns (300 ns) for send (receive), overlapped between the 12 independent
  directions and an aggregate bandwidth of 12 Gbit/s
\item other components such as Ethernet controller, interrupt
  controller, etc.
\end{itemize}
The chip will occupy a die size of about $(12 {\rm mm})^2$, and the
power consumption is expected to be in the range of 1--2 W.

Two such ASICs will be mounted on a daughterboard, together with two
industry-standard double data rate (DDR) SDRAM modules (one per ASIC)
whose capacity will be determined by the price of memory when the
machine is assembled.  (At the time of this writing, 256 MByte
registered DIMMs with ECC were available for \$40.)  32 daughterboards
will be mounted on a motherboard, and 8 motherboards in a crate with a
single backplane.  The final machine then consists of a certain number
of such crates.

There are two separate networks: the physics network and an
Ethernet-based auxiliary network.  The physics network consists of
high-speed serial links between nearest neighbors with a bandwidth of
2$\times$500 Mbits/s per link.  The nodes are arranged in a
6-dimensional torus which allows for an efficient partitioning of the
machine in software, as described in detail in Ref.~\cite{Chen01}.
The serial-communications unit in the QCDOC ASIC provides direct
memory access, single-bit error detection with automatic resend, and a
low-latency store-and-forward mode for global operations.  The auxiliary
network is used for booting, diagnostics, and I/O over Ethernet, with
an Ethernet controller integrated on the ASIC.  The Ethernet traffic
to and from the ASIC will run at 100 Mbit/s.  Hubs or switches on the
motherboard will provide a bandwidth of 1.6 Gbit/s off a motherboard
to commercial switches and the host workstation.

\section{THE QCDOC ASIC}

\begin{figure*}[!t]
%  \centerline{\epsfig{figure=asic.eps,width=100mm}}
\vskip -5mm \hskip -40mm
  \epsfig{figure=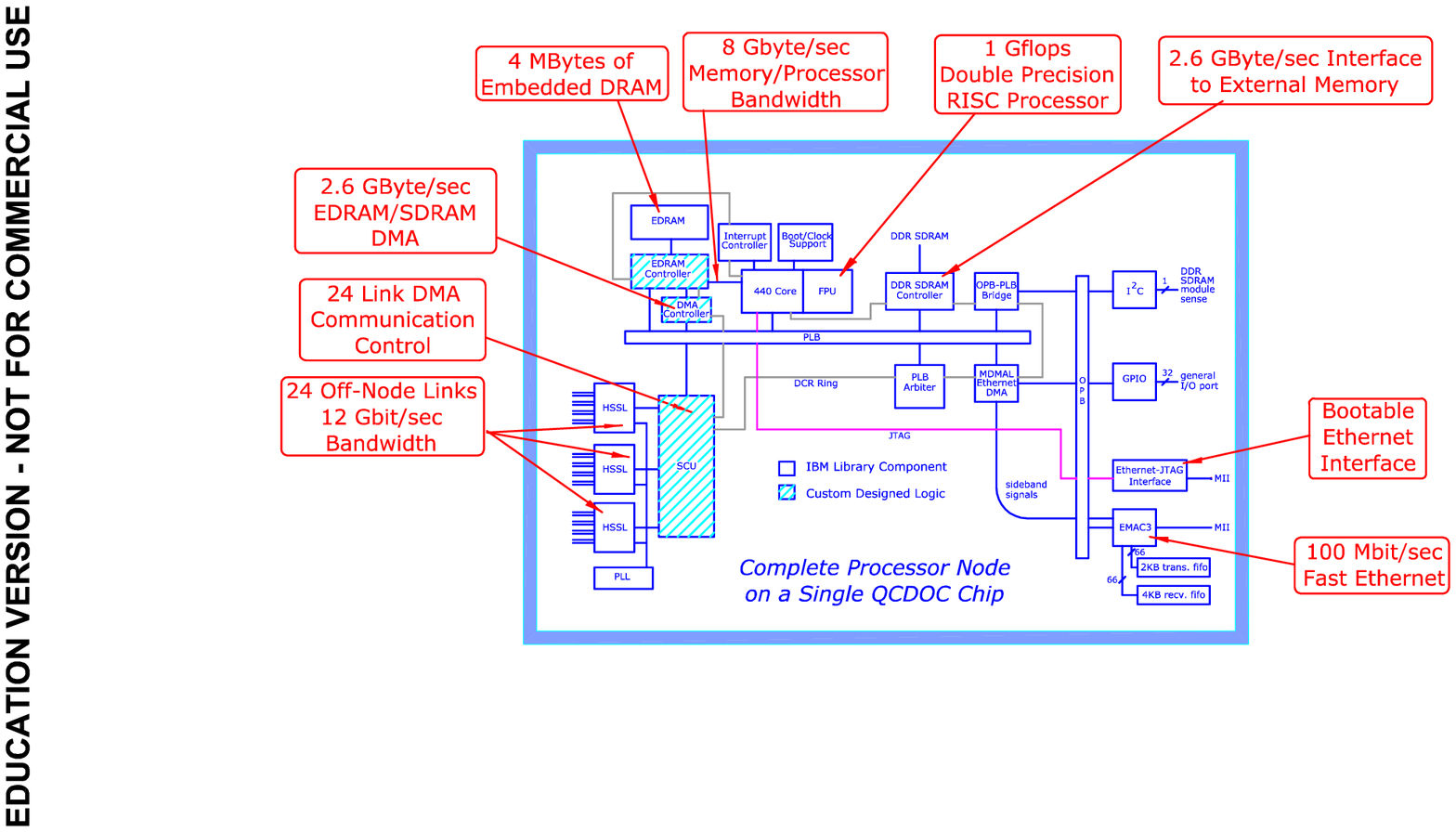,width=200mm}
  \vspace*{-32mm}
  \caption{The QCDOC ASIC.  The cross-hatched parts represent
    custom-designed logic.} 
  \label{fig:asic}
\end{figure*}

A schematic picture of the QCDOC ASIC is shown in Fig.~\ref{fig:asic}.
In this section we describe the various components in more detail.
Many of the ingredients of the ASIC are provided by IBM, see
Secs.~\ref{sec:ibmlib}--\ref{sec:ejtag}.  The major custom-designed
components are the serial communications unit (SCU) and the
prefetching eDRAM controller (PEC), see
Secs.~\ref{sec:scu}--\ref{sec:pec}.

\subsection{IBM Library Components}
\label{sec:ibmlib}

\medskip

\begin{itemize}\itemsep -1mm
\item There are three main busses:
  \begin{itemize}
  \item The processor local bus (PLB) is a 128-bit wide, fully
    synchronous bus that runs at 1/3 of the CPU frequency.  It has
    separate read and write busses, supports pipelining (up to 4
    deep), and has a large number of sophisticated features such as
    split transactions, burst transfers, etc.  
  \item The on-chip peripheral bus (OPB) is a 32-bit wide, synchronous
    bus running at 1/2 of the PLB frequency.  Its basic purpose
    is to off-load slower devices from the PLB bus.
  \item The device control register bus (DCR) is a slow and simple bus
    used to read and write control registers.  The devices on this bus
    are connected in a ring-like fashion.  The 440 is the only master
    on this bus.
  \end{itemize}
\item The 440 PowerPC processor core is a dual-issue, superscalar,
  32-bit implementation of the IBM Book-E architecture (the E stands
  for embedded).  The core has a 7 stage pipeline.  There
  are instruction and data caches of size 32 kBytes each.  These are
  64-way associative, partitionable, and lockable.  Hardware memory
  protection is provided through a translation-lookaside buffer (TLB).
  The 440 has three PLB master interfaces, one each for instruction
  read, data read, and data write.  The 440 will run at
  500 MHz and is connected to a 64-bit, 1 GFlops IEEE floating-point
  unit (for details of the FPU see Ref.~\cite{Dockser01}).  An
  important feature of the 440 is that it contains a JTAG interface.
  JTAG (Joint Test Action Group) is an industry-standard protocol that
  allows an external device to take complete control of the processor.
  This functionality will be used for booting and debugging, see
  below.
\item The PLB arbiter provides programmable arbitration for the up to
  eight allowed masters that can control PLB transfers.  In our design
  we have seven masters: the three PLB master interfaces of the 440 (two
  of which, namely data read and data write, will be channeled
  through the PEC, see below), the MCMAL, two master interfaces for the
  send and receive operations of the SCU and one for the DMA unit.
\item The 4 MBytes of embedded DRAM (eDRAM) are a standard IBM library
  component that will be accessed with low latency and high bandwidth
  through a custom-designed controller, the PEC, see
  Sec.~\ref{sec:pec} below.
\item The Universal Interrupt Controller (UIC) processes the various
  interrupts generated on- and off-chip and provides critical and
  non-critical interrupt signals to the 440.
\item The DDR controller is a slave on the PLB, capable of transferring
  data to/from external DDR SDRAM at a peak bandwidth of 2.6 GBytes/s.
  It supports an address space of 2 GBytes and provides error
  detection, error correction, and refresh of the off-chip SDRAM.
\item The PLB-OPB bridge is used to transfer data between the two
  busses.  It is the only master on the OPB and a slave on the PLB.
\item The Ethernet media access controller (EMAC) provides a 100 Mbit/s 
  Ethernet interface.
  The media-independent interface (MII) signals at the ASIC boundary
  are connected to a physical layer chip on the daughterboard.  The
  EMAC is a slave on the OPB and has sideband signals to the MCMAL on
  the PLB.
\item The DMA-capable Memory Access Layer (MCMAL) 
  loads/unloads the EMAC through the sideband signals.  It is a master
  on the PLB.
\item The inter-integrated circuit (I$^2$C) controller is a slave on
  the OPB used to read the configuration of the DDR DIMM.  This
  information is then used to configure the DDR controller.
\item The general purpose I/O (GPIO) unit is another slave on the OPB
  whose 32-bit wide data bus will be taken out to the ASIC boundary.
  It can be used, e.g., to drive LEDs.
\item The high-speed serial links (HSSL) used by the SCU each contain
  eight independent ports (four each for send/receive) through which
  bits are clocked into/out of the ASIC at 500 Mbit/s per port.  The
  bits are converted to bytes (or vice versa) in the HSSL.  The HSSL
  input clocks are phase-aligned by another IBM macro, the phase-locked loop
  (PLL).
\end{itemize}

\subsection{Ethernet-JTAG Interface}
\label{sec:ejtag}

As mentioned above, the 440 core has a JTAG interface over which one
can take complete control of the processor.  In particular, this
interface can be used to load boot code into the instruction cache and
start execution.  This completely eliminates the need for boot ROM.
The question is how the JTAG instructions should be loaded into the
440.  (There are special tools that use the JTAG interface, but it
would be impractical to connect one tool per ASIC for booting.)  A
solution to this question has already been developed at IBM Research,
implemented using a field-programmable gate array (FPGA), that converts
special Ethernet packets to JTAG commands and vice versa.  This logic
is now part of the QCDOC ASIC and will be used not only for booting
but also to access the CPU for diagnostics/debugging at run time.
The unique MAC address of each ASIC is provided to the Ethernet-JTAG
component by location pins on the ASIC, and the IP address is then
derived from the MAC address.

\subsection{Serial Communications Unit (SCU)}
\label{sec:scu}

\begin{figure*}[!t]
  \begin{center}
    \epsfig{figure=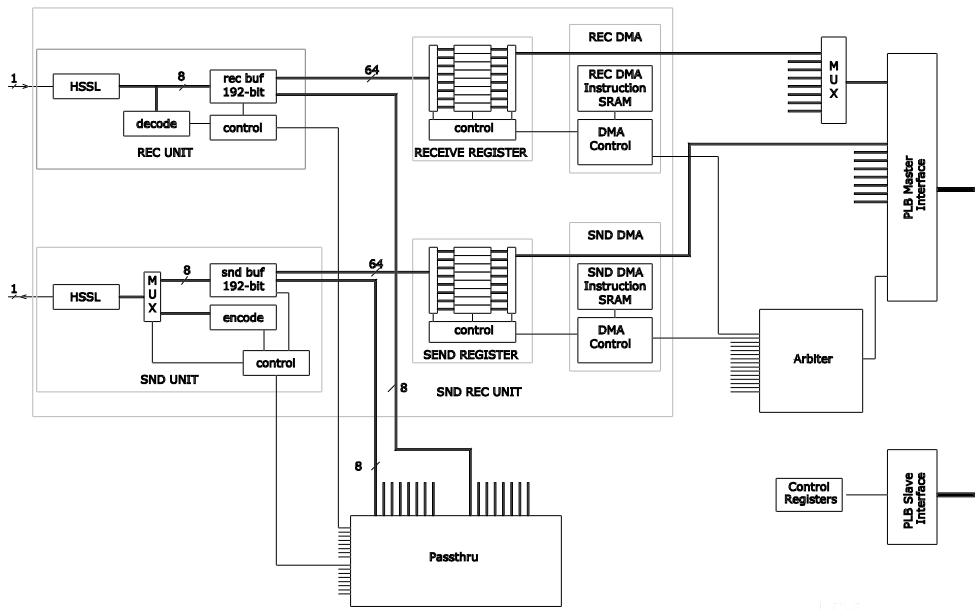,width=80mm,angle=270}
    \vspace*{-3mm}
    \caption{Serial Communications Unit in the QCDOC ASIC.  The ASIC
      boundary is on the left, the PLB interface on the right.}
    \label{fig:scu}
  \end{center}
\end{figure*}

The task of the SCU is to reliably manage the exchange of data between
neighboring nodes with minimum latency and maximum bandwidth.  The
design takes into account the particular communication requirements of
lattice QCD simulations.  A schematic picture of the SCU is shown in
Fig.~\ref{fig:scu}.

The custom protocol governing the data transfers defines packets that
contain a 64-bit data word and an 8-bit header containing control and
parity bits.  When the receive unit receives a packet, it first
interprets the header, buffers the bytes from the HSSL, and assembles
the 64-bit word.  It then transfers the word to the receive register
or passes it on to the send unit.  The receive buffer can store three
64-bit words so that the send unit (in a neighboring node) can send
three words before an acknowledgment is received.  The functionality
of the send unit is essentially the inverse of that of the receive
unit.  Send and receive operations can proceed simultaneously.  
Each send or receive unit is controlled by a DMA engine which then 
transfers the data between memory and a send/receive register.  
Each DMA engine is controlled by block-strided-move instructions stored 
in SRAM in the SCU itself.

A low-latency passthrough mode is provided for global operations.
Because of the latencies associated with the HSSL, the most efficient
method to perform global sums is ``shift-and-add'', using
a store-and-forward capability built into the SCU. 

The expected latency of a send (receive) operation is 120 ns (300 ns).  This is
one to two orders of magnitude lower than the latencies associated
with Myrinet.  Since a write instruction from the 440 can initiate many
independent transfers on any subset of the 24 send or receive channels, 
the latencies associated with multiple transfers can be overlapped to 
some degree.  The total off-chip bandwidth using all 24 HSSL ports is
12 Gbit/s.  In a 4-dimensional physics application only 16 of the 24
HSSL ports will be used, resulting in a total bandwidth of 8 Gbit/s.
This provides a good match for the communications requirements of
typical applications.  As a worst-case example, we consider a $2^4$
local lattice and the application of a staggered fermion conjugate gradient.
In this case we obtain

\medskip

\noindent\begin{tabular}{@{\hspace*{3mm}}l@{\hspace*{5mm}}l}
  FPU alone & 75\% efficiency\\
  FPU + n.n.\ communications & 48\% efficiency\\
  FPU + n.n.\ + global sums & 38\% efficiency
\end{tabular}

\medskip

\noindent The 75\% efficiency in the first line is a reasonable
assumption based on the performance figures quoted in
Table~1 below.

\subsection{\mbox{Prefetching eDRAM Controller (PEC)}}
\label{sec:pec}

\begin{figure}[!t]
  \begin{center}
    \epsfig{figure=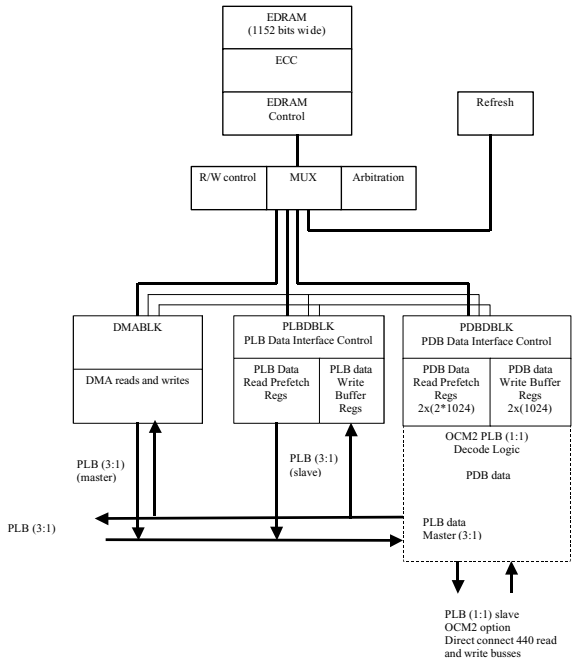,width=\columnwidth}
    \vspace*{-10mm}
    \caption{Prefetching eDRAM Controller in the QCDOC ASIC.}
    \label{fig:pec}
  \end{center}
\vskip -10mm
\end{figure} 

The PEC (shown schematically in Fig.~\ref{fig:pec}) is designed to
provide the 440 with high-bandwidth access to the eDRAM. 
It interfaces to the 440 data read and data write busses via a fast
version of the PLB that runs at the CPU frequency and that we call
processor direct bus (PDB).
The PEC also contains a PLB slave interface to allow for read
and write operations from/to any master on the PLB as well as a DMA
engine to
transfer data between eDRAM and the external DDR memory.

The maximum PDB bandwidth is 8 GBytes/s for read and 8 GBytes/s for
write.  The latency of the PDB itself is 1-2 CPU cycles.  This very
low latency eliminates the need for pipelining.
The access to eDRAM (which is memory-mapped) proceeds at 8
GBytes/s.  ECC is built into the PEC, with 1-bit error correct and
2-bit error detect functionality.  The PEC also refreshes the eDRAM.

The read data prefetch from eDRAM occurs in two 1024-bit lines.  Three
read ports (PDB, PLB slave, DMA) arbitrate for the common eDRAM.  The
coherency between PDB, PLB slave, and DMA is maintained within the
PEC.  Each read port has four 1024-bit registers that are paired in
two sets to allow for ping-ponging between different memory locations.
There are also two 1024-bit write buffer registers each for the
PDB/PLB slave/DMA write interfaces.

\section{SOFTWARE}

In this section the various aspects of the \mbox{QCDOC} software
environment are described.  Efforts are being made to make this
environment user-friendly and to allow a wide range of application
codes to run on the QCDOC platform without the need for extensive
porting.

{\bf Boot kernel}.  The boot kernel will be loaded from the host work
station via Ethernet using the reliable UDP protocol.  The
Ethernet-JTAG interface in the ASIC converts the packets to JTAG
commands which in turn load the boot code into the instruction cache
and start execution (see also Sec.~\ref{sec:ejtag}).  The boot kernel
then assigns an IP address to the ASIC (based on the MAC address
provided by the location pins) and loads the Ethernet driver.

{\bf Run-time kernel}.  The run-time kernel communicates with the host
using the Ethernet driver loaded by the boot kernel.  It performs
complete node diagnostics and provides read, write, and execute
capability.  It is based on the UNIX RPC protocol.

{\bf RISCWatch debugger}.  This is an IBM-provided PowerPC
development/debugging environment for a set of selected nodes.  It
provides single-stepping capability with an advanced, user-friendly
graphical user interface showing a cycle-by-cycle display of the
internal state of the PowerPC processor as OS or user code executes.

{\bf Extensive diagnostic programs}.  These will be provided to permit
extensive, real-time hardware monitoring during program execution as
well as targeted diagnostic tests for hardware maintenance. 

{\bf Host Operating System}.  The host machine will be an SMP
workstation with a suitable number of Gbit Ethernet ports, using a
threaded transport protocol to efficiently manage tens of thousands of
nodes.  The host OS will manage both the boot and the run-time kernel
operation.  It will provide both a qcsh and a socket-based interface
to the QCDOC operation, thus allowing for both csh-scripted and
Perl-based machine control.

{\bf Disk software}.  We are planning a parallel disk system
supporting a general distribution of disks.  A simplified UNIX-like
hierarchical file system will be used.  A utility will be provided to
store parallel files as a single, flattened, position-indexed file for
archiving and use on other systems.

{\bf Compiler, linker, loader}.  The compiler and linker are GNU-based tools
supporting C and C++ as well as Book-E compliant, PowerPC assembly
language.

{\bf Lattice QCD application code}.  A code structure is being developed 
that will both provide the cross-platform capabilities needed by the UKQCD
Collaboration and follow the QCD-API standard being developed by the
U.S.\ SciDAC-supported lattice QCD software initiative.

\section{STATUS AND PERFORMANCE\\ FIGURES}
\label{sec:perf}

The functional design of the QCDOC ASIC is nearly finished.  Single-node
physics code is running on the VHDL simulator, and from these
simulations we obtained the performance figures shown in Table~1. 
These all refer to double-precision calculations.
\begin{table*}[!t]
  \label{tab:perf}
  \begin{center}
    Table 1.  Double-precision performance of single-node physics code running
    on the simulator.\\[3mm]
    \begin{tabular}{l@{\hspace*{20mm}}l}
      \hline\\[-1mm]
      single node FPU, SU(3) $\times$ 2-spinor & 84\% of peak from L1
      cache\\ & 78\% of peak from eDRAM\\
      cache fill/flush to eDRAM & 8.0 GB/s peak\\
      & 3.2 GB/s sustained (due to 440 latency)\\
      software visible bandwidth to eDRAM &  2.5 GB/s read,
      2.0 GB/s write\\ 
      DDR fill/flush & 1.2 GB/s read, 1.7 GB/s write\\[2mm]
      \hline
    \end{tabular}
  \end{center}
\end{table*}

\begin{figure}[!t]
  \begin{center}
    \epsfig{figure=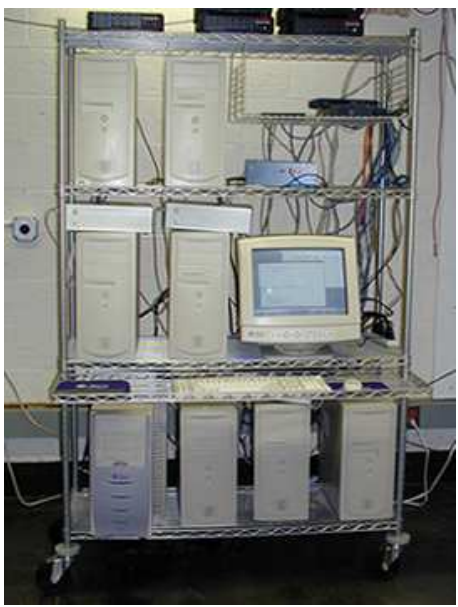,width=60mm}
    \vspace*{-10mm}
    \caption{Multi-processor software development platform with prototype 
             Ethernet/JTAG hardware.}
    \label{fig:ppc_devel}
  \end{center}
\vskip -10mm
\end{figure}

We have set up a test environment in which two nodes can successfully
communicate in the VHDL simulator over their Ethernet interfaces
and/or the HSSLs.  The operating system is being developed on a system
with six PowerPC 405GP boards, one PowerPC 750 board, five
Ethernet-JTAG boards, and a RISCWatch probe.  This hardware configuration 
is shown in Figure~\ref{fig:ppc_devel}.

The current design has been successfully synthesized, and an analysis
net list has been provided to IBM for floorplanning purposes.
Anticipating a few iterations to satisfy timing constraints 
the complete design should be transferred to IBM
by the end of 2001.

\section{CONCLUSIONS}

The QCDOC architecture combines state-of-the-art system-on-a-chip
technology with custom-designed logic specially optimized for lattice
QCD calculations to provide computing power in the tens of TFlops
range at a price/performance ratio of 1~US-\$ per sustained MFlops.
Optimized QCD code is expected to run with an efficiency of roughly
50\%.  The communications latencies of the QCDOC design are one to
two orders of magnitude smaller than those of commercial network
solutions used in PC clusters, thus allowing for a large problem to be
distributed onto many nodes.  This fact, along with significantly
lower power consumption, higher reliability, more compact packaging and a better
price/performance ratio, motivates the design efforts that are required
to construct such a supercomputer.

The first QCDOC machines with a combined peak speed of about 7 TFlops
are projected to be finished in 2002, and by the end of 2003 the
combined peak speed of QCDOC machines world-wide is targeted at 40
TFlops (10 TFlops of which will be installed in Edinburgh, 10 TFlops are 
planned for the RBRC and 20 TFlops for the U.S.\ lattice community).

\section*{ACKNOWLEDGMENTS}

This research was supported in part by the U.S. Department of Energy,
the Institute of Physical and Chemical Research (RIKEN) of Japan, and
the U.K. Particle Physics and Astronomy Research Council.


\begin{thebibliography}{1}
\bibitem{Luescher01} M. L\"uscher, hep-lat/0110007, these
  proceedings. 
\bibitem{Chen01} D. Chen et al., Nucl. Phys. B (Proc. Suppl.) 94
  (2001) 825.
\bibitem{Dockser01} {\tt www.chips.ibm.com/products/powerpc/} {\tt
  newsletter/aug2001/new-prod3.html} 
\end{thebibliography}
\end{document}